\documentclass[%
 reprint,
superscriptaddress,
 amsmath,amssymb,
prl,
]{revtex4-2}

\usepackage{graphicx}
\usepackage{dcolumn}
\usepackage{bm}

\begin{document}

\preprint{APS/123-QED}

\title{A telecom O-band emitter in diamond}

\author{Sounak Mukherjee}
\altaffiliation{These authors contributed equally to this work.}
\author{Zi-Huai Zhang}
\altaffiliation{These authors contributed equally to this work.}
\affiliation{Department of Electrical and Computer Engineering, Princeton University, Princeton, New Jersey 08544, USA}

\author{Daniel G. Oblinsky}
\affiliation{Department of Chemistry, Princeton University, Princeton, New Jersey 08544, USA}

\author{Mitchell O. de Vries}
\affiliation{School of Science, RMIT University, Melbourne, Victoria 3000, Australia}

\author{Brett C. Johnson}
\affiliation{ARC Centre of Excellence for Quantum Computation and Communication Technology, School of Engineering, RMIT University, Melbourne, Victoria 3001, Australia}

\author{Brant C. Gibson}
\affiliation{ARC Centre of Excellence for Nanoscale BioPhotonics, School of Science, RMIT University, Melbourne, Victoria 3001, Australia}
\author{Edwin L. H. Mayes}
\affiliation{RMIT Microscopy and Microanalysis Facility, RMIT University, Melbourne, Victoria 3001, Australia}
\author{Andrew M. Edmonds}
\author{Nicola Palmer}
\author{Matthew L. Markham}
\affiliation{Element Six, Harwell, OX11 0QR, UK}

\author{\'Ad\'am Gali}
\affiliation{Wigner Research Centre for Physics, P.O. Box 49, 1525 Budapest, Hungary}
\affiliation{Department of Atomic Physics, Institute of Physics, Budapest University of Technology and Economics, M\H{u}egyetem rakpart 3., 1111 Budapest, Hungary}
\author{Gerg\H{o} Thiering}
\affiliation{Wigner Research Centre for Physics, P.O. Box 49, 1525 Budapest, Hungary}

\author{Adam Dalis}
\author{Timothy Dumm}
\affiliation{Hyperion Materials \& Technologies, 6325 Huntley Road, Columbus, Ohio 43229, USA}

\author{Gregory D. Scholes}
\affiliation{Department of Chemistry, Princeton University, Princeton, New Jersey 08544, USA}

\author{Alastair Stacey}
\affiliation{School of Science, RMIT University, Melbourne, Victoria 3000, Australia}

\author{Philipp Reineck}
\affiliation{ARC Centre of Excellence for Nanoscale BioPhotonics, School of Science, RMIT University, Melbourne, Victoria 3001, Australia}

\author{Nathalie P. de Leon}
\email{npdeleon@princeton.edu }
\affiliation{Department of Electrical and Computer Engineering, Princeton University, Princeton, New Jersey 08544, USA}

\date{\today}

\begin{abstract}
Color centers in diamond are promising platforms for quantum technologies. Most color centers in diamond discovered thus far emit in the visible or near-infrared wavelength range, which are incompatible with long-distance fiber communication and unfavorable for imaging in biological tissues. Here, we report the experimental observation of a new color center that emits in the telecom O-band, which we observe in silicon-doped bulk single crystal diamonds and microdiamonds. Combining absorption and photoluminescence measurements, we identify a zero-phonon line at 1221~nm and phonon replicas separated by 42~meV. Using transient absorption spectroscopy, we measure an excited state lifetime of around 270~ps and observe a long-lived baseline that may arise from intersystem crossing to another spin manifold.
\end{abstract}
\maketitle

Quantum emitters are widely explored for quantum technologies \cite{o2009photonic,buluta2011natural,azzam2021prospects, de2021materials}, particularly quantum communication and quantum sensing applications \cite{atature2018material, awschalom2018quantum}. For quantum communication, transmission over long distances requires the use of existing fiber-based optical communication networks \cite{cao2019telecom} where losses are high in the visible and near-infrared region. For comparison, visible light can undergo a fiber loss of 8~dB/km, whereas it is about 0.5 dB/km at 1300~nm and 0.2~dB/km at 1550~nm \cite{agrawal2012fiber}. For sensing applications, the second near-infrared (NIR-II) window spanning 1000-1700 nm is particularly attractive for biological imaging because of deeper optical penetration, lower absorption and higher contrast due to reduced  background autofluorescence and scattering from biological tissues \cite{duan2018recent}. 

Some of the most widely studied quantum emitters are color centers in diamond \cite{ruf2021quantum,pezzagna2021quantum,rodgers2021materials}, namely the nitrogen vacancy (NV$^-$) \cite{maletinsky2012robust,bernien2013heralded}, negative silicon vacancy (SiV$^-$) \cite{rogers2014all,pingault2017coherent, nguyen2019integrated, nguyen2019quantum} and neutral silicon vacancy (SiV$^0$) \cite{rose2018observation,zhang2020optically} centers, all of which emit photons in the visible to near-infrared region. To date, a few defects in other host materials have been reported that emit in the telecom band: the G-, T- and W-centers in silicon \cite{udvarhelyi2021identification,safonov1996interstitial, baron2022detection}, NV$^-$, vanadium, and molybdenum in silicon carbide \cite{zargaleh2018nitrogen, wolfowicz2020vanadium, bosma2018identification}, and erbium in various host materials \cite{yin2013optical,alizadehkhaledi2019isolating,raha2020optical,dibos2018atomic,cajzl2017erbium}, but no color centers in diamond have been reported in the telecom band. Diamond offers several unique advantages as a host -- it has a large optical transparency window, has the highest Debye temperature of any material, is chemically inert and biocompatible, and can provide a low magnetic noise environment with the low natural abundance of $^{13}$C \cite{pezzagna2021quantum}. Hence, there is significant motivation to search for telecom band emitters in diamond.

Here, we report the experimental observation of an emitter in diamond that emits at 1221 nm, which is near the `original' wavelength band (O-band) used for optical communication. Using photoluminescence (PL) and absorption spectroscopy, we identify a zero-phonon line (ZPL) at 1221~nm and phonon replicas spaced by 42~meV. Using transient absorption spectroscopy, we probe the excited state lifetime of the defect and also observe a long-lived shelving state. We tentatively assign this defect as a silicon-related defect, and we compare it to prior theoretical reports \cite{thiering2015complexes}.

The samples studied in this work are silicon doped samples with different form factors and synthesis conditions, from bulk diamonds to microdiamonds and nanodiamonds. The three bulk diamonds (labelled samples 1, 2, and 3) were grown by plasma chemical vapor deposition (CVD, Element Six) and doped with silicon during growth \cite{Supplemental}. These contain ensembles of SiV$^0$, and optical and spin characterization of SiV$^0$ in sample 1 were reported in Ref.~\onlinecite{rose2018strongly,zhang2020optically}. Silicon-doped microdiamonds were made using high-pressure high-temperature (HPHT) synthesis (Hyperion Materials \& Technologies), followed by bead milling to reduce the size of the as-synthesized crystals. Nanodiamond particles were extracted from this sample using centrifugation \cite{Supplemental}. The microdiamonds have been reported to exhibit fluorescence from silicon-related defects including SiV$^-$ and silicon-boron color centers \cite{shames2020near}, and we also observe PL from SiV$^0$ (Fig.~S6 \cite{Supplemental}). In our study, broadband absorption measurement was conducted using a UV-Vis-NIR spectrophotometer (Agilent, Cary 5000) at room temperature. Low temperature optical measurements were performed in home-built confocal microscopes equipped with 850~nm optical excitation and a spectrometer with a liquid nitrogen cooled InGaAs detector (PyLoN IR, Princeton Instruments). Transient absorption measurements were conducted using a femtosecond, broadband transient absorption spectrometer \cite{pensack2016observation} (Helios, Ultrafast Systems) with the sample cooled to 4.2~K. 
\begin{figure}[t!]
  \centering
  \includegraphics[width = 86mm]{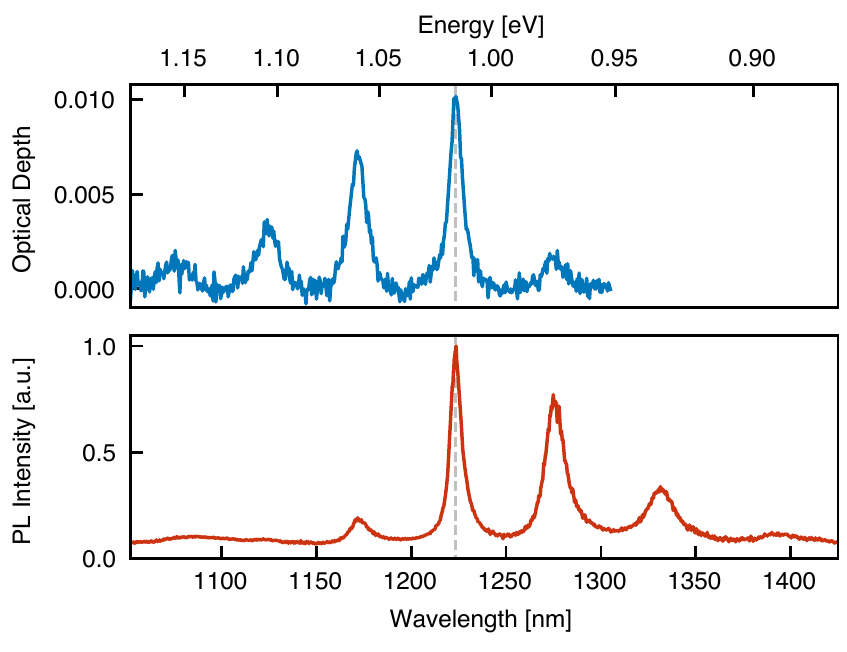}
  \caption{Absorption and PL spectroscopy. Room temperature absorption (top) and PL (bottom) spectra at 285~K for sample 1 with narrow peaks evenly spaced by approximately 42~meV. The two spectra display a clear mirror symmetry, with the mirror axis being the brightest line at 1224~nm. The grey dashed line shows the axis of symmetry as a guide to the eye. The thickness of the sample is 0.66 mm.}
  \label{fig:Fig1}
\end{figure}

First, we probe the telecom band emitter by correlating absorption and PL measurements on a bulk CVD diamond, sample 1, at room temperature (Fig.~\ref{fig:Fig1}). Using 850~nm excitation, we observe several PL peaks at 1173~nm, 1224~nm, 1276~nm, 1332~nm, and 1393~nm that are evenly spaced and separated by approximately 42~meV. Absorption spectroscopy reveals similar evenly spaced peaks at 1076~nm, 1124~nm, 1172~nm, 1224~nm, and 1275~nm, consistent with prior observations of absorption lines in these bands in silicon doped diamond \cite{johansson2011optical}. The absorption and PL spectra display a clear mirror symmetry with the symmetry axis being the most intense peak at 1224~nm (1.0129 eV). The symmetry between absorption and emission suggests the peaks originate from the same defect, with a ZPL at 1224~nm. 

\begin{figure}[b!]
  \centering
  \includegraphics[width = 86mm]{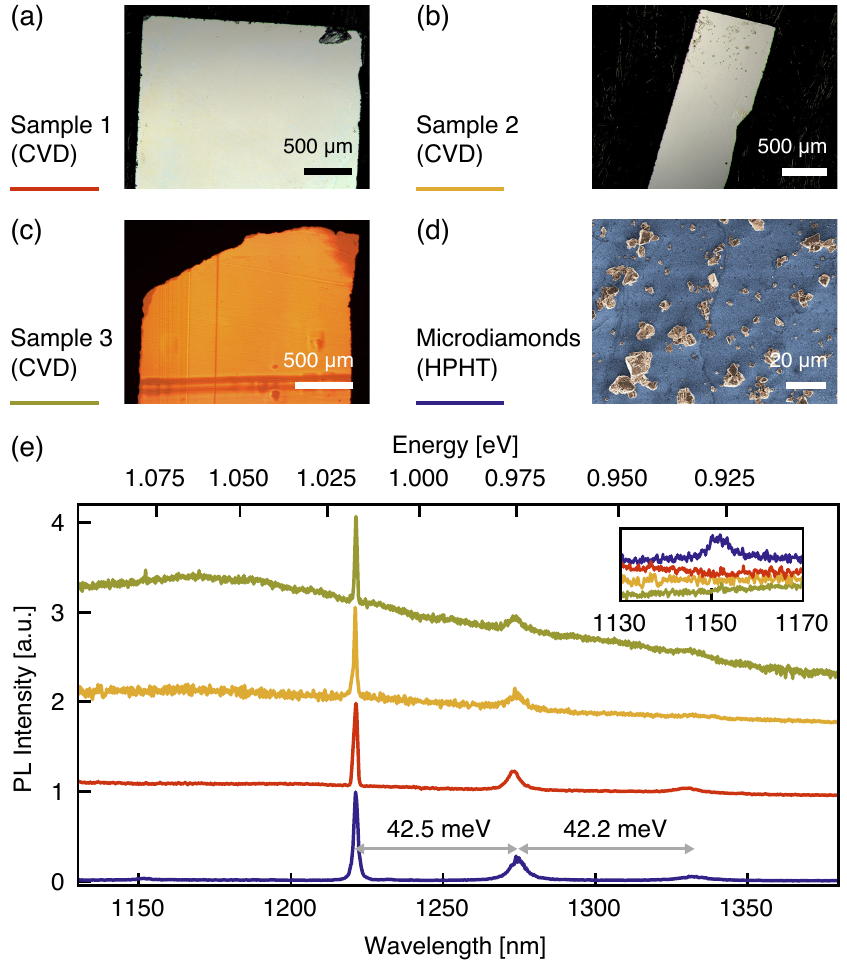}
  \caption{Low temperature PL in different samples. Optical images of (a) sample 1, (b) sample 2, and (c) sample 3, which are bulk-doped diamonds grown by CVD. (d) False-colored scanning electron microscope image of HPHT microdiamonds. (e) Comparison of the PL emission spectra in different samples with 850~nm excitation. Color coding of the spectra corresponds to samples shown in (a) to (d). Inset shows PL intensity from 1130~nm to 1170~nm where we observe an additional peak at 1152~nm for the microdiamonds. The data in the inset has been rescaled to the same background level with an arbitrary offset to separate the spectra. The PL spectra corresponding to samples 1 and 3 have been measured at 8.5~K. For sample 2 and the microdiamonds, the temperature was 5.8~K.}
  \label{fig:Fig2}
\end{figure}

With the assignment of the absorption and emission peaks to a common defect, we probe its optical properties at cryogenic temperatures. We measure the emission spectrum from 1100~nm to 1400~nm using 850~nm excitation on the three bulk silicon doped samples (Fig.~\ref{fig:Fig2}(a), (b) and (c)) and silicon doped microdiamonds (Fig.~\ref{fig:Fig2}(d)). Despite their drastically different form factors, synthesis conditions, and preparation methods, we observe three evenly spaced PL peaks (1221~nm, 1273~nm, and 1330~nm) common to all samples~(Fig.~\ref{fig:Fig2}(e)). The peak at 1221~nm shows the narrowest linewidth among the three peaks, consistent with it being the ZPL. Additional features such as a broad background in samples 2 and 3 and an additional peak at 1152~nm in the microdiamonds are observable, but they do not correlate across samples, indicating they are not related to the same defect. We can also observe the 1221~nm transition in silicon doped sub-100 nm nanodiamonds, although with a much lower emission intensity (Fig.~S2 \cite{Supplemental}). The spacing (42~meV) between the peaks is consistent with theoretical calculations of the localized vibronic mode arising from the vibration of a silicon atom in various silicon containing complexes \cite{thiering2015complexes}. Based on this observation, together with the abundance of silicon in these samples, we hypothesize that the 1221~nm emission arises from a silicon-related defect. We note that all the diamonds in which we have observed the telecom emitter also contain SiV$^0$, which may suggest that its charge transition energy is near the valence band maximum. The microscopic nature of this defect requires further investigation but absorption lines in this band were proposed to arise from SiV$_2$:H$^{(-)}$ in a recent theoretical study \cite{thiering2015complexes}.

\begin{figure}[h!]
  \centering
  \includegraphics[width = 86mm]{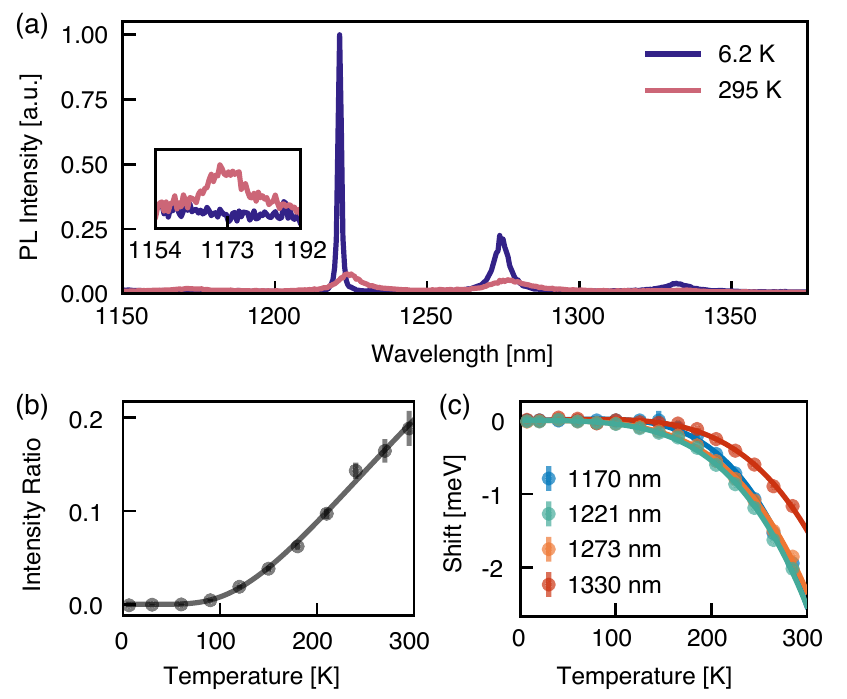}
  \caption{Temperature dependence of PL. (a) Emission spectra of microdiamonds measured at 6.2~K and room temperature (295~K). Inset shows the spectra around 1170~nm, where an additional peak appears at room temperature. Data points in the spectra have been averaged with four points in each bin. (b) Temperature dependence of the ratio of the intensity of the 1170~nm peak to the 1221~nm peak in microdiamonds, fitted to $Ae^{-\Delta E/k_B T}$ with an activation energy of $41.8\pm1.7$~meV. (c) Temperature dependence of the wavelength shift of each of the PL lines in sample 1, fitted to $\mu T^4 + \nu T^2$ \cite{Supplemental}.}
  \label{fig:Fig3}
\end{figure}

Comparing the room temperature and low temperature spectra, we observe that the emission peak at 1170~nm is present at room temperature but not at low temperature~(Fig.~\ref{fig:Fig3}(a)). Hence, we investigate the optical properties of the telecom emitter by studying the temperature dependence of the PL spectra in detail. At temperatures higher than 90 K, the peak at 1170~nm becomes visible (Fig.~S3(a)). The ratio of the intensity of the 1170~nm peak compared to the ZPL follows an exponential temperature dependence (Fig.~\ref{fig:Fig3}(b)). By fitting the ratio of intensities with $Ae^{-\Delta E/k_B T}$, we obtain an activation energy of $41.8\pm1.7$~meV, consistent with the separation between the 1170~nm peak and the ZPL (44.9~meV at 295~K). The occurrence of this higher energy peak and its exponential activation with temperature are consistent across multiple samples (microdiamonds: Fig.~3(b), sample 1: Fig.~S3(c)). Therefore, we attribute this 1170~nm peak to the emission of a photon from a vibrational excited state in the electronic excited state manifold. We note that we do not observe a uniform trend for the temperature dependence of the total PL intensity across different samples, which may be related to defect-defect interactions and charge state dynamics. 

At higher temperatures, the PL peak positions shift to longer wavelengths and the linewidths broaden. For comparison, the linewidth at 6.2~K for the microdiamonds is $1.12\pm0.01$~nm, whereas the room temperature linewidth is $8.76\pm0.14$~nm. The magnitude of the shift is similar to previous reports for NV \cite{hizhnyakov2002zero, davies1974vibronic} and SiV$^-$ \cite{neu2013low} centers, where the shifts were attributed to the expansion of the diamond lattice along with temperature-dependent electron-phonon interactions. The volume expansion coefficient of diamond has been empirically shown to have a cubic dependence on temperature \cite{stoupin2011ultraprecise}, which causes the lattice expansion to contribute a $T^4$ term to the energy shift. This, combined with the softening of elastic bonds in the excited states \cite{hizhnyakov2002zero}, leads to a $\mu T^4 + \nu T^2$ dependence (sample 1: Fig. \ref{fig:Fig3}(c), microdiamonds: Fig. S3(b) \cite{Supplemental}).

\begin{figure}
  \centering
  \includegraphics[width = 86mm]{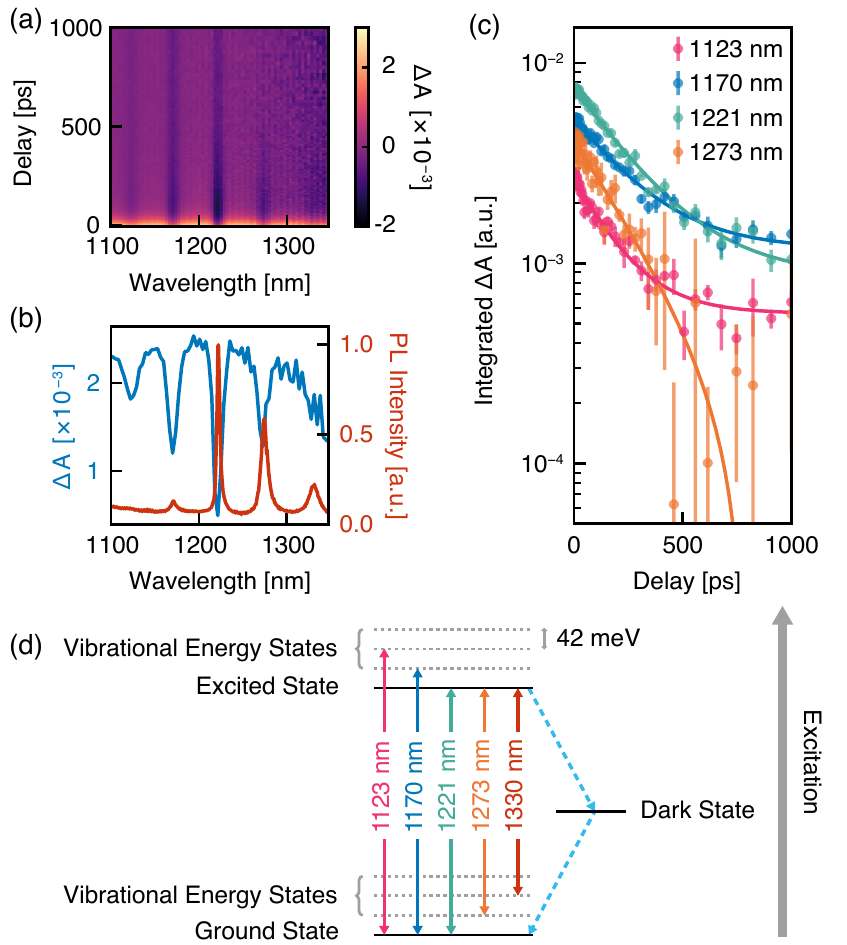}
  \caption{Transient absorption spectroscopy. (a) 2D map of transient absorption ($\Delta A$) as functions of time delay and wavelength measured at 4.2~K. A series of approximately evenly spaced lines in wavelength originating from the telecom emitter are observable. Wavelengths are shifted by 2.9~meV to adjust for spectrometer calibration. (b) Comparison between the PL spectrum (red) and the transient absorption spectrum (blue). The respective peak and trough positions occur at the same wavelengths. The PL spectrum is taken at 225~K. The transient absorption spectrum is a linecut from (a) at a delay of 5~ps. (c) Decay of integrated $\Delta A$ for different transitions. Each point is an average of four data points, with error bars given by standard error. The decays are fitted to a single exponential and we observe decay constants of $177.8\pm13.9$~ps$, 256.1\pm16.1$~ps, $261.4\pm12.8$~ps, and $272.1\pm37.8$~ps,  for the 1123~nm, 1170~nm, 1221~nm, and 1273~nm transitions respectively. (d) Level diagram of the telecom emitter deduced from PL emission, absorption, and transient absorption. Here the wavelength labels correspond to their values measured at low temperature.}
  \label{fig:Fig4}
\end{figure}

To study the optical dynamics of the defect, we perform pump-probe measurements using transient absorption spectroscopy on sample 1. We apply a femtosecond 937 nm pump pulse (FWHM 16~nm, $1.5\times10^{15}$~photons/cm$^2$, Fig.~S1 \cite{Supplemental}) to excite the ground state population and use a broadband (820-1350~nm) probe pulse to monitor the change in absorption after the pump. The repetition rate for the sequence is kept at 1~kHz to ensure full relaxation of the population in the beginning of each sequence. The difference in absorption, $\Delta A$, with and without the pump is measured at different pump-probe time separations (Fig.~\ref{fig:Fig4}(a)). Within the $\Delta A$ spectrum, we observe minima that are consistent with the steady-state absorption spectrum (Fig.~\ref{fig:Fig1}) and match with the peak positions in the PL spectra (Fig.~\ref{fig:Fig4}(b)). 

The nature of these minima depends on their relative position to the ZPL. The higher energy dips (1123~nm, 1170~nm) arise from absorption from the ground state to different vibrational states in the excited state. They appear as minima in the $\Delta A$ spectrum (Fig.~\ref{fig:Fig4}(b)) because the ground state population is bleached by the pump, leading to less absorption of the probe pulse. The appearance of the lower energy dip at 1273~nm is inconsistent with the absorption process at 4.2 K, as it would require the thermal population of a vibrationally excited mode (42~meV) in the ground state, which is a much larger energy than the thermal energy ($\sim$0.36~meV). Instead, this feature at 1273~nm is likely the result of stimulated emission \cite{berera2009ultrafast}, where the excited state is first populated after the pump and then emission is stimulated with the probe, creating a dip in $\Delta A$. Similarly, the $\Delta A$ dip at the ZPL can have contributions from both ground state bleach and  stimulated emission. 

The temporal dependence of $\Delta A$ reveals the kinetics of the population dynamics. We observe exponential decays for the integrated $\Delta A$ at each band (Fig.~\ref{fig:Fig4}(c)). The decay constant for the different peaks are similar (around 270~ps), suggesting they share the same decay paths. The decay of $\Delta A$ for the 1273~nm dip is attributed to the decay of the excited state population. Therefore, we infer the lifetime of this emitter to be $272.1\pm37.8$~ps. We note that sample 1 also shows signs of charge state conversion between the telecom emitter and SiV$^0$ (Fig.~S6 \cite{Supplemental}), which may affect the measured lifetime in this sample. We observe different baselines for the different $\Delta A$ bands. The decay for the 1273~nm transition reaches a value of zero, suggesting all the population from the excited state is relaxed. However, the curves for 1123~nm, 1170~nm and 1221~nm decay to a non-zero baseline, suggesting that the ground state population is trapped in a dark shelving state with a lifetime much longer than the duration of the measurement (1~ns). 

Combining PL, absorption, and transient absorption, we propose a tentative level structure for this telecom emitter (Fig.~\ref{fig:Fig4}(d)). The emission band at 1221~nm is assigned to the ZPL because it is the mirror axis of absorption and emission spectra. The defect has a silicon-related vibrational mode with an energy of 42~meV. The population of this mode in the excited state leads to the anti-Stokes lines at 1170~nm and 1123~nm, while the population of this mode in the ground state leads to the Stokes lines at 1273~nm and 1330~nm. Finally, we propose the existence of an additional dark state with longer lifetime coupled to the optically active states through an intersystem crossing. We note that this level structure resembles the level structure of its proposed SiV$_2$:H$^{(-)}$ assignment \cite{thiering2015complexes}. For SiV$_2$:H$^{(-)}$, the level structure was proposed to have a singlet-to-singlet optical transition with a triplet shelving state. We carried out additional calculations on this model which further strengthen the tentative assignment of the center to the SiV$_2$:H$^{(-)}$ defect model. We find that the short PL lifetime originates from the fast non-radiative decay from the singlet excited state to the singlet ground state~\cite{Supplemental}.

To summarize, we have experimentally observed telecom O-band PL emission from a silicon-related defect in CVD and HPHT diamonds. A tentative level structure is proposed based on PL emission, absorption, and transient absorption. More experimental investigations are required to confirm its microscopic structure and symmetry. In order to confirm the defect structure, silicon ion implantation followed by hydrogen diffusion and annealing could be used to form silicon containing centers decorated with hydrogen \cite{glover2003hydrogen,fuchs1995hydrogen}. In addition to longer propagation lengths in optical fiber, the telecom band emission wavelength leads to reduced scattering in nanophotonic devices \cite{burek2014high}, allowing for more efficient integration for long distance quantum networks. Furthermore, diamond nanoparticles hosting the telecom emitter can be potentially incorporated as bioimaging probes by virtue of their form factor and emission in the NIR-II window.

This work was primarily supported by the U.S.
Department of Energy, Office of Science, National Quantum Information Science Research Centers, Co-design
Center for Quantum Advantage (C2QA) under contract
number DE-SC0012704. Ultrafast measurements and materials characterization were performed with support from the Air Force Office of Scientific Research under Grant No. FA9550-17-0158. This research was also supported by the Australian Research Council Center of Excellence for Nanoscale BioPhotonics (No. CE140100003) and the Australian Research Council Center of Excellence for Quantum Computation and Communication Technology (No. CE170100012).  P.R. acknowledges funding through the RMIT Vice-Chancellor's Research Fellowship and ARC DECRA Fellowship scheme (No. DE200100279). M.dV. acknowledges funding through the RMIT's Research Stipend Scholarship (RRSS-SC). The authors acknowledge the use of the RMIT Microscopy and Microanalysis Facility (RMMF) at RMIT University. \'A.~G.~acknowledges the support from the NKFIH in Hungary for the National Excellence Program (Grant No.\ KKP129866), the Quantum Information National Laboratory, and the EU QuantERA II Sensextreme project. G.~T.~was supported by the János Bolyai Research Scholarship of the Hungarian Academy of Sciences. G.~T.~acknowledges the high-performance computational resources provided by KIF\"U (Governmental Agency for IT Development) institute of Hungary. 

\bibliography{TelecomEmitter}
\end{document}


\title{Supplemental Material for \\``A telecom O-band emitter in diamond''}

\author{Sounak Mukherjee}
\altaffiliation{These authors contributed equally to this work.}
\author{Zi-Huai Zhang}
\altaffiliation{These authors contributed equally to this work.}
\affiliation{Department of Electrical and Computer Engineering, Princeton University, Princeton, New Jersey 08544, USA}

\author{Daniel G. Oblinsky}
\affiliation{Department of Chemistry, Princeton University, Princeton, New Jersey 08544, USA}

\author{Mitchell O. de Vries}
\affiliation{School of Science, RMIT University, Melbourne, Victoria 3000, Australia}

\author{Brett C. Johnson}
\affiliation{ARC Centre of Excellence for Quantum Computation and Communication Technology, School of Engineering, RMIT University, Melbourne, Victoria 3001, Australia}

\author{Brant C. Gibson}
\affiliation{ARC Centre of Excellence for Nanoscale BioPhotonics, School of Science, RMIT University, Melbourne, Victoria 3001, Australia}
\author{Edwin L. H. Mayes}
\affiliation{RMIT Microscopy and Microanalysis Facility, RMIT University, Melbourne, Victoria 3001, Australia}

\author{Andrew M. Edmonds}
\author{Nicola Palmer}
\author{Matthew L. Markham}
\affiliation{Element Six, Harwell, OX11 0QR, UK}

\author{\'Ad\'am Gali}
\affiliation{Wigner Research Centre for Physics, P.O. Box 49, 1525 Budapest, Hungary}
\affiliation{Department of Atomic Physics, Institute of Physics, Budapest University of Technology and Economics, M\H{u}egyetem rakpart 3., 1111 Budapest, Hungary}
\author{Gerg\H{o} Thiering}
\affiliation{Wigner Research Centre for Physics, P.O. Box 49, 1525 Budapest, Hungary}

\author{Adam Dalis}
\author{Timothy Dumm}
\affiliation{Hyperion Materials \& Technologies, 6325 Huntley Road, Columbus, Ohio 43229, USA}

\author{Gregory D. Scholes}
\affiliation{Department of Chemistry, Princeton University, Princeton, New Jersey 08544, USA}

\author{Alastair Stacey}
\affiliation{School of Science, RMIT University, Melbourne, Victoria 3000, Australia}

\author{Philipp Reineck}
\affiliation{ARC Centre of Excellence for Nanoscale BioPhotonics, School of Science, RMIT University, Melbourne, Victoria 3001, Australia}

\author{Nathalie P. de Leon}
\email{npdeleon@princeton.edu }
\affiliation{Department of Electrical and Computer Engineering, Princeton University, Princeton, New Jersey 08544, USA}

\date{\today}
\maketitle
\label{Sec:SI}

\setcounter{figure}{0}
\setcounter{section}{0}
\renewcommand{\thefigure}{S\arabic{figure}}
\renewcommand{\thetable}{\Roman{table}}
\renewcommand{\thesection}{\Roman{section}}
\renewcommand{\theequation}{S\arabic{equation}}

\section{\label{SI_Setup} EXPERIMENTAL METHODS}
\subsection{Experimental Setup}
Photoluminescence (PL) spectroscopy was performed in two home-built confocal microscopes with different imaging modalities. 

The first confocal microscope is optimized for bulk PL detection and is used for samples 1 and 3.
The sample is cooled inside either a helium vapor cryostat (Oxford OptistatCF) or a helium flow cryostat (Janis ST-100). A near-infrared coated lens is used to focus the excitation onto the sample, with the emission being collected into a 50~$\mu$m multimode fiber. The excitation and emission are separated using a dichroic beamsplitter (Semrock FF925-Di01-25x36). An 850~nm fiber-coupled diode laser (QPhotonics QFLD-850-100S) is used for excitation. The excitation is filtered with a 900~nm short pass filter while the detection is filtered with a 900~nm long pass filter. The collected PL emission is routed to a spectrometer (Princeton Instruments SpectraPro HRS-300) equipped with a liquid nitrogen cooled InGaAs detector (PyLoN IR). For resonant excitation of neutral silicon vacancy (SiV$^0$) centers, the dichroic beamsplitter is replaced with a 50:50 beamsplitter (Thorlabs BSW26R), while the excitation source is replaced with a tunable diode laser (Toptica CTL 950). The detection is filtered with a 980~nm long pass filter (Semrock LP02-980RE-25).

The second confocal microscope is optimized for high numerical aperture (NA) imaging, which we use to study the emission from sample 2 and microdiamonds and sub-100~nm nanodiamonds. The sample is cooled inside a close-cycle cryostat (Montana S50) equipped with a 0.65 NA objective (Olympus LCPLN50NIR) inside vacuum. Confocal raster scans are performed using a dual-axis scanning galvo (ThorLabs GVS012). An 850~nm fiber-coupled diode laser (QPhotonics QFLD-850-100S) is used for excitation. A polarizer (Thorlabs LPVIS050-MP2) and a half-wave plate (Thorlabs AHWP10M-980) are used to control the polarization of the excitation. The excitation and detection channels are combined with a dichroic beam splitter (Semrock FF925-Di01-25x36). 
The detection channel is filtered with an 1150~nm long pass filter (Thorlabs FELH1150). The collected PL emission is split into two paths using a 75:25 fiber beam splitter (Thorlabs TW930R3A1). One of the paths is routed to a superconducting nanowire single-photon detector (SNSPD), while the other path is routed to a spectrometer (Princeton Instruments SpectraPro HRS-300) equipped with a liquid nitrogen cooled InGaAs detector (PyLoN IR).

\begin{figure}[h!]
  \centering
  \includegraphics[width = 136mm]{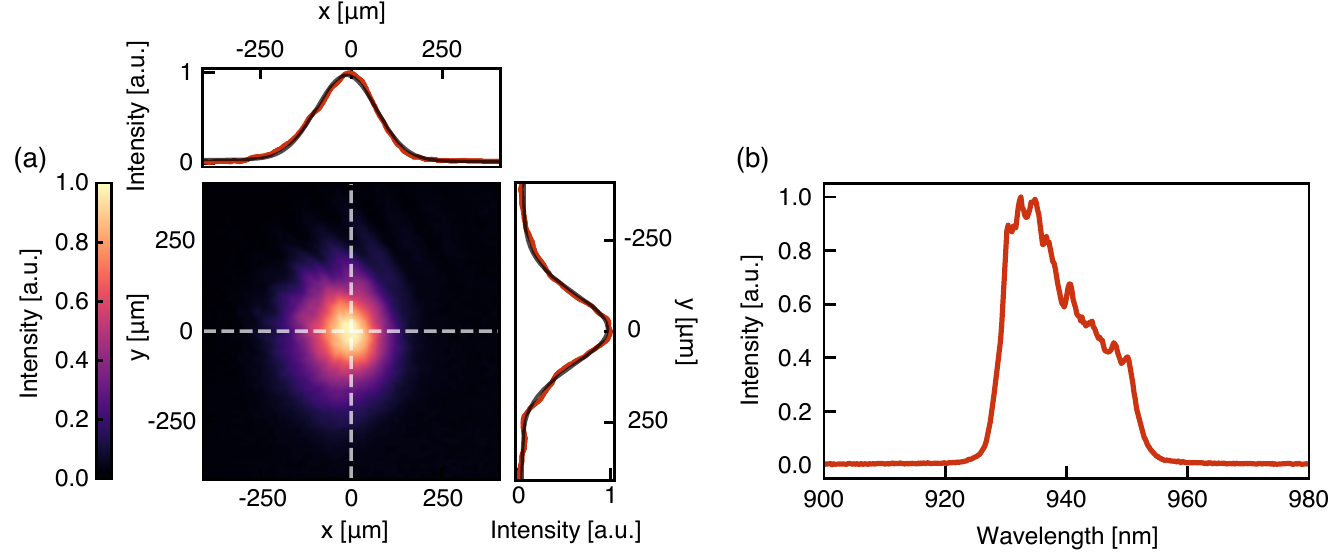}
  \caption{(a) Excitation beam profile for transient absorption spectroscopy taken at sample position. The line profiles at the positions marked by white dashed lines are plotted on the top and right (red). Black curves are Gaussian fits. Full width at half maximum (FWHM) along $x$ and $y$ directions are found to be $194.6\pm1.1$~$\mu$m and $233.2\pm1.4$~$\mu$m respectively. Pixel size is 4.65 $\mu$m. (b) Spectrum of the excitation beam used for transient absorption spectroscopy.}
  \label{fig:TA_beam}
\end{figure}

Transient absorption measurements were conducted using a femtosecond, broadband transient absorption spectrometer (Helios, Ultrafast Systems) at the ultrafast laser spectroscopy facility in the Frick Chemistry Laboratory at Princeton University. The details of the spectrometer were described in Ref.~\onlinecite{pensack2016observation}. For our measurements, the sample was cooled to 4.2~K inside a helium flow cryostat. The full width at half maximum for the pump pulse was approximately 200 $\mu$m at the sample location (Fig.~\ref{fig:TA_beam}(a)). The pump pulse was filtered to center around 937~nm with a spectral bandwidth of 16~nm (Fig.~\ref{fig:TA_beam}(b)).

\subsection{Sample Preparation}
Sample 1 and sample 2 are \{110\} oriented diamonds grown by plasma chemical vapor deposition (Element Six). Sample 3 was originally grown on a \{100\} substrate, but subsequently processed into a \{110\} cross-section.  Samples 1 and 3 were doped with $^{29}$Si during growth. Sample 1 contains SiV$^0$ centers around 200 ppb, and was previously characterized in Ref.~\onlinecite{rose2018strongly,zhang2020optically}. Sample 2 was doped with $^{28}$Si during growth, and contains approximately 30 ppb of SiV$^0$ centers based on absorption measurements.

The high pressure high temperature (HPHT) silicon doped diamond particles \cite{shames2020near} are commercially available from Ad\'amas Nanotechnologies. They are synthesized by Hyperion M\&T and are $\sim1\mu$m in size (Fig. 2(d)). To obtain the sub-100~nm nanodiamonds, the microdiamond particle powder was suspended in deionized water (10 mg/mL), sonicated for 30 min (100 W, 66\% duty cycle) and then centrifuged to separate the smallest particles. The suspension was centrifuged twice (1000 rcf, 10 minutes), with the larger particles in the centrifugate discarded and the supernatant used for further experiments. The transmission electron microscope image of the nanoparticles are shown in Fig.~\ref{fig:ND_100nm}(a). Intensity weighted distribution from dynamic light scattering of the nanoparticles suspended in water shows that most of their dimensions lie below 100~nm (Fig.~\ref{fig:ND_100nm}(b)). Before coating the particles onto a substrate, we break down the large aggregates using ultrasonication and disperse the diamond particles in deionized water. The microdiamonds are then dropcasted on a quartz chip for PL spectroscopy. The sub-100~nm nanodiamonds are spin coated for a sparser distribution on a sapphire substrate. 

\section{\label{SI_EXP} SUPPLEMENTARY MEASUREMENTS}
\subsection{Telecom emission in sub-100~nm diamond particles}
\begin{figure}[h!]
  \centering
  \includegraphics[width = 172mm]{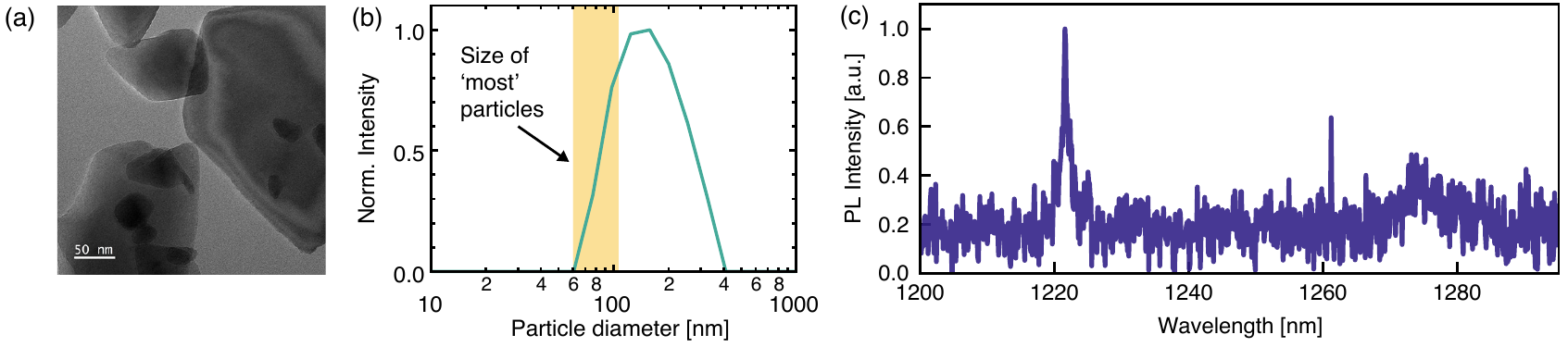}
  \caption{(a) Transmission electron microscope image of sub-100~nm nanodimonds. (b) Intensity weighted distribution of the dimensions of sub-100~nm nanodiamonds from dynamic light scattering. (c) PL spectrum of the sub-100~nm nanodimonds at 5.8 K showing the ZPL (1221~nm) and phonon replica (1273~nm) from the telecom emitter.
}
  \label{fig:ND_100nm}
\end{figure}

The PL spectrum from sub-100~nm diamond particles is shown in Fig.~\ref{fig:ND_100nm}(c). Similar to the microdiamonds and the bulk-doped samples, the 1221~nm zero phonon line (ZPL) is observed together with the phonon replica at 1273~nm. The emission intensity in the sub-100~nm diamond particles is much lower compared to the bulk-doped samples and microdiamonds (Fig. 2(a)), likely due to the lower number of emitters, higher effective sample temperature and broadening in the nanoparticles. 

\subsection{Temperature dependence of telecom emission}
\begin{figure}[h!]
  \centering
  \includegraphics[width = 150mm]{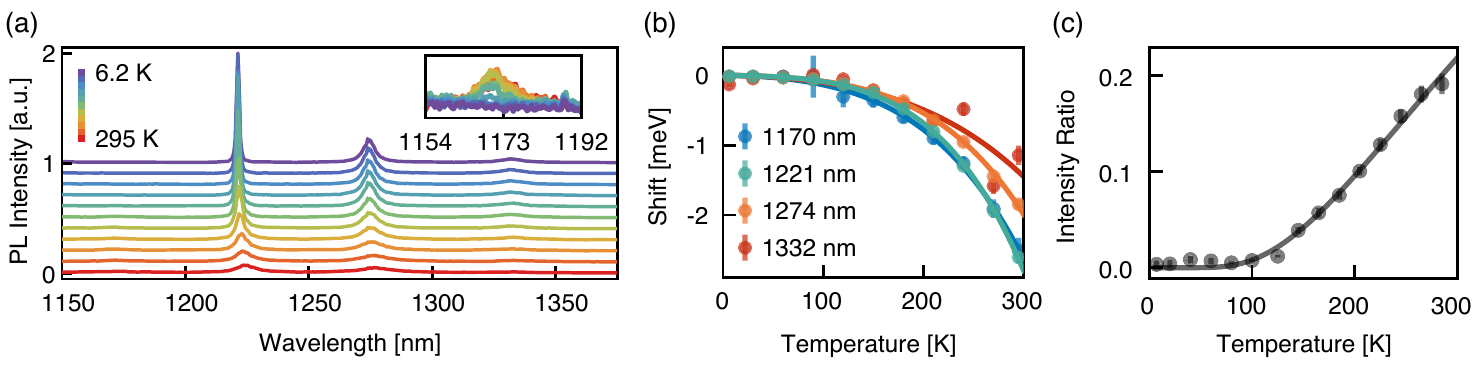}
  \caption{(a) Temperature dependent PL emission spectra for microdiamonds from 6.2~K to room temperature. Inset: zoom-in spectra from 1154~nm to 1192~nm. The $\sim$1170~nm peak is visible at temperatures higher than 90~K. (b) Shift of different transitions with temperature in microdiamonds, fitted to $\mu T^4 + \nu T^2$. (c) Ratio of 1170~nm peak intensity to ZPL intensity as a function of temperature in sample 1. The data is fitted using $Ae^{-\Delta E/k_B T}$ (solid line), with an activation energy of $42.8\pm1.9$~meV.}
  \label{fig:temp_dep}
\end{figure}

The PL emission spectra of microdiamonds at 6.2~K and room temperature are shown in Fig. 3(a). The full range of temperature-dependent spectra are plotted in Fig.~\ref{fig:temp_dep}(a). The inset shows the occurrence of the phonon replica at 1170~nm at high temperatures. The energy shift corresponding to the different transitions of the telecom emitter in microdiamonds are shown in Fig. \ref{fig:temp_dep}(b). Fitting the shift with $\mu T^4 + \nu T^2$ gives us $\mu = -(2.79 \pm 0.09)\times10^{-10}$~meV~K$^{-4}$ and $\nu = -(5.99 \pm 0.76)\times 10^{-6}$~meV~K$^{-2}$ for the ZPL. This is consistent with the temperature-dependent shift for sample 1 shown in Fig. 3(c), where the fitting parameters for ZPL are obtained as $\mu = -(2.87 \pm 0.14)\times10^{-10}$~meV~K$^{-4}$ and $\nu = -(2.17 \pm 1.04)\times 10^{-6}$~meV~K$^{-2}$. The shift of the telecom emitter transitions can be useful for temperature sensing in biological applications. From the ZPL shift in sample 1 near room temperature, we can estimate the temperature accuracy (2 standard deviations) as 0.8~K for 30 seconds of integration time, giving rise to a sensitivity of 4.2~K~Hz$^{-1/2}$.

The intensity ratio of the 1170~nm phonon replica to the ZPL for sample 1 is plotted in Fig.~\ref{fig:temp_dep}(c). For temperatures below 145~K, the 1170~nm peak intensity is calculated by integrating the peak area under the curve to prevent poor fitting. We get an activation energy of $42.8\pm1.9$~meV by fitting the data with $Ae^{-\Delta E/k_B T}$. This is in good agreement to the temperature dependence for microdiamonds (Fig. 3(b)).

\subsection{Polarization and power dependence of the telecom emission}
\begin{figure}[h!]
  \centering
  \includegraphics[width = 86mm]{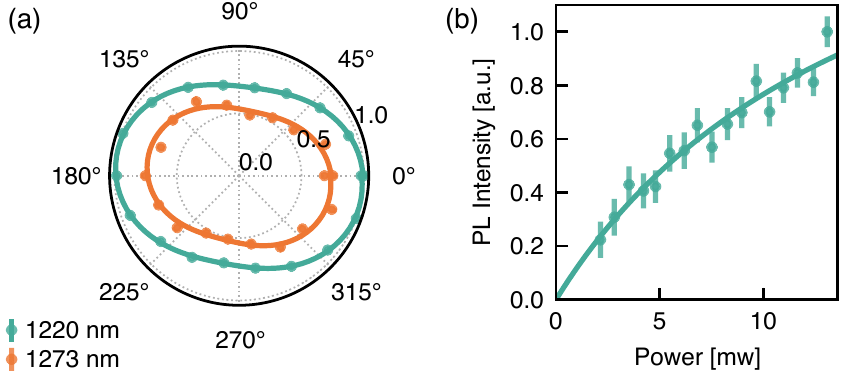}
  \caption{(a) Excitation polarization dependence of PL intensity for the two transitions from the telecom emitter in microdiamonds. Solid lines are sinusoidal fits to the data. (b) ZPL intensity as a function of  excitation power for sub-100~nm nanodiamonds. At high powers, the dependence deviates from a linear scaling. The data is fit to the saturation form $I_\infty P/(P+P_\text{sat})$, with a saturation power ($P_\text{sat}$) as  $15.7\pm4.0$~mW.}
  \label{fig:pol_pow}
\end{figure}

We probe the telecom emission in microdiamonds by measuring the emission intensity with respect to the polarization of the excitation. For most of the particles, the emission shows no polarization dependence, as expected from ensemble averaging of a large number of emitters along different crystal axis. In some particles, both the ZPL (1220~nm) and the phonon replica (1273~nm) intensities show a sinusoidal variation with the excitation polarization (Fig.~\ref{fig:pol_pow}(a)).

By measuring the PL intensity as a function of excitation power for the microdiamonds showing strong polarization dependence, we observe PL saturation at high powers. This is more pronounced for the sub-100~nm nanodiamonds, whose saturation curve for the ZPL is shown in Fig.~\ref{fig:pol_pow}(b). We note that a linear scaling of the power dependence is expected for large ensembles. Observing PL saturation, together with a strong polarization dependence, indicates that some of the sub-100~nm nanodiamonds contain a small number of telecom emitters.

\subsection{Time-resolved PL decay of telecom emission}
Time-resolved PL decay measurements were carried out at 10~K with a custom-built scanning confocal fluorescence microscope and a closed cycle liquid helium cryostat (Cryostation, Montana Instruments, USA). A pulsed 532~nm laser (pulse-width $<$85~ps, 5~MHz repetition rate, VisUV-532, PicoQuant, Germany) was used for excitation and focused using a 50x air objective (NA~0.65, LCPLN50xIR, Olympus, Japan). The fluorescence signal was separated from the excitation beam using a 532~nm dichroic beam splitter (LPD02-532RU-25, Semrock, USA), a 532~nm long-pass filter (BLP01-532R-25, Semrock, USA) and a 1200~nm long-pass filter (Edmund Optics, USA). An avalanche photodiode (ID230, IDQuantique, Switzerland) was employed for imaging and time-resolved fluorescence decay measurements and a spectrometer (IsoPlane320 with a PyLoN-IR: 1024-1.7 camera, Princeton Instruments, USA) to obtain fluorescence spectra. A correlator card (TimeHarp 260, Picoquant, Germany) was used to analyze photon arrival times to create time-resolved direct fluorescence decay traces.
\begin{figure}[h!]
  \centering
  \includegraphics[width = 66mm]{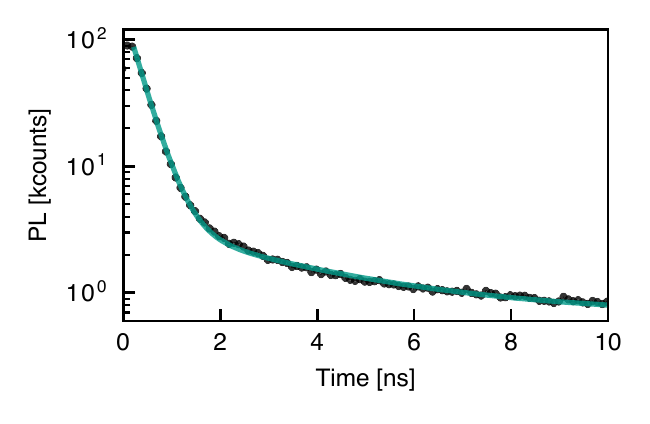}
  \caption{Time-resolved PL decay from microdiamonds. Black dots are experimental data and the teal solid line is a double exponential fit. This gives us a lifetime of $322.6\pm2.5$~ps for the telecom emitter.}
  \label{fig:PL_decay}
\end{figure}

Fig. \ref{fig:PL_decay} shows the time-resolved PL decay from the microdiamonds, where all the emission above 1200~nm is collected. Fitting to a double exponential form gives us a short timescale of $322.6\pm2.5$~ps, which is an estimate for the lifetime of the telecom emitter. The longer decay constant of $3.2\pm0.6$~ns is hypothesized to be related to the emission from the tail of an unknown defect with $\sim$900~nm ZPL, as reported in Ref.~\onlinecite{shames2020near}. The fitted amplitudes indicate that about 98\% of the counts are from the emitter with the shorter lifetime. Although the accuracy of this measurement is limited by the response time of the system, it is in good agreement with the $\sim270$~ps lifetime measured from transient absorption.

\subsection{Observation of neutral silicon vacancy centers in microdiamonds}
\begin{figure}[h!]
  \centering
  \includegraphics[width = 172mm]{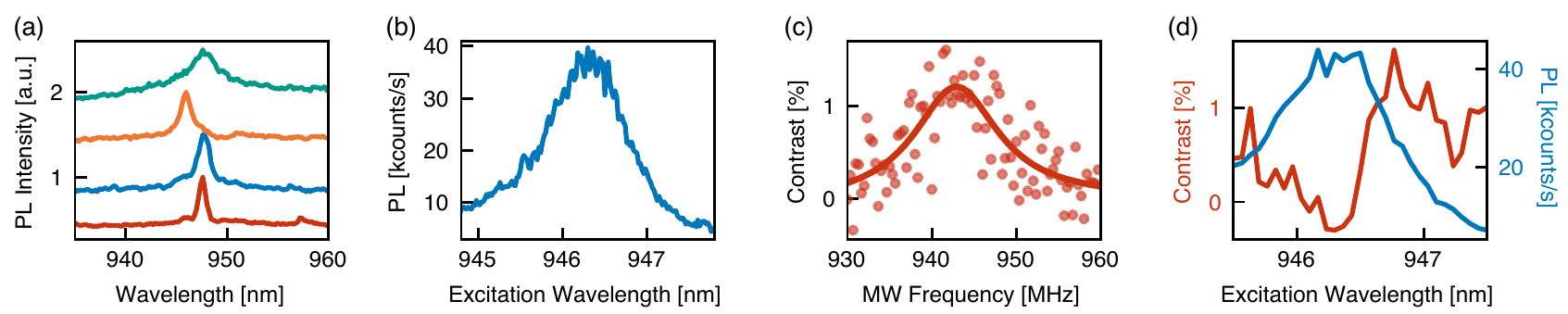}
  \caption{(a) PL spectrum of different microdiamonds at 8~K under 850~nm excitation showing the ZPL for SiV$^0$ at $\sim946$~nm. (b) Photoluminescence excitation (PLE) spectrum of SiV$^0$ in microdiamonds. (c) Continuous-wave optically detected magnetic resonance (ODMR) with Lorentzian fit. (d) Wavelength dependence of ODMR contrast (red) with PLE (blue) on the same microdiamond as reference.}
  \label{fig:SiV0_in_ND}
\end{figure}

Previous studies have reported the presence of SiV$^0$ in the microdiamonds from electron paramagnetic resonance \cite{shames2020near}. By performing PL measurements on these microdiamonds, we also observe the PL emission of SiV$^0$, where a narrow line around the ZPL of SiV$^0$ (946~nm) is visible, as shown in Fig.~\ref{fig:SiV0_in_ND}(a). The SiV$^0$ centers in the microdiamonds are stable and allow for resonant excitation via the ZPL~(Fig. \ref{fig:SiV0_in_ND}(b)). We can further observe optically detected magnetic resonance (ODMR) of SiV$^0$ in the microdiamonds (Fig. \ref{fig:SiV0_in_ND}(c)) as we sweep the microwave (MW) frequency. Lorentzian fit gives us a zero field splitting of $942.9\pm0.5$~MHz. With the microwave on resonance, we can observe the wavelength dependence of ODMR contrast (Fig. \ref{fig:SiV0_in_ND}(d)). This is consistent with measurements on SiV$^0$ in bulk-doped samples.

\subsection{Enhancement of the telecom emission by resonant excitation of SiV$^0$ in sample 1}
\begin{figure}[h!]
  \centering
  \includegraphics[width = 86mm]{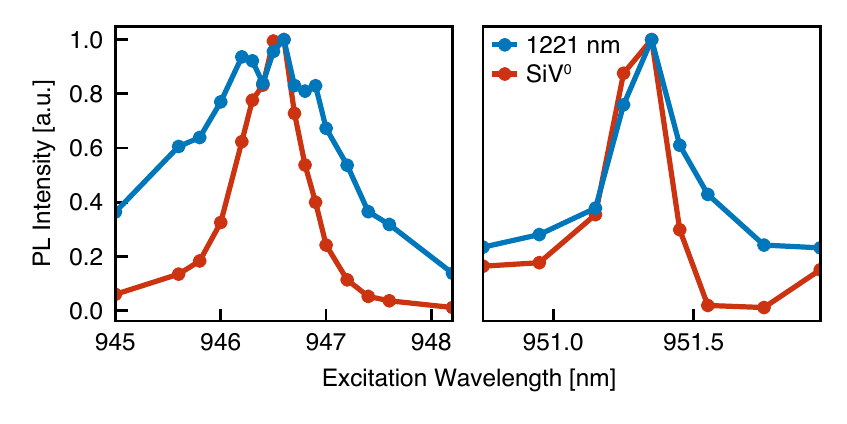}
  \caption{PL intensity of both the telecom emitter and SiV$^0$ as a function of excitation wavelength in sample 1. Both the SiV$^0$ sideband emission (red) and 1221~nm peak of the telecom emitter (blue) are maximized at SiV$^0$ ZPL. The intensity is calculated by integrating the area under the spectrum.}
  \label{fig:PLE}
\end{figure}

We observe non-trivial enhancement of the telecom emission via resonant excitation of SiV$^0$ in sample 1. As the excitation sweeps across SiV$^0$ ZPL at 946~nm and 951~nm, we observe a strong enhancement of the ZPL of the telecom emitter (Fig.~\ref{fig:PLE}). This enhancement is only observed in sample 1, which has the highest concentration for both SiV$^0$ and the telecom emitter. Therefore, we hypothesize that the enhancement may arise from charge transfer process between the two different emitters via charge state conversion of SiV$^0$ through resonant excitation.

We note that the transient absorption measurements were performed on sample 1 with the pump pulse exciting SiV$^0$ resonantly. Therefore, the excited state lifetime of $\sim$270~ps extracted using transient absorption measurement may be complicated by the charge state dynamics of SiV$^0$ and the telecom emitter in this sample.

\section{\label{SI_Setup}Theoretical description of $\text{SiV}_2$:H$^{(-)}$ model}

In this section, we discuss the SiV$_2$:H$^{(-)}$ model of the 1221~nm PL center. In our previous study~\cite{thiering2015complexes}, we tentatively assigned the 1.018~eV absorption center to the SiV$_2$:H$^{(-)}$ defect based on the calculated ZPL energy and sharp phonon replica that we associated with the motion of Si ion in the defect. We also showed in that study that hydrogen vibration is not involved in the optical transition. We carry out further calculations to determine the PL lifetime of the defect, in order to further strengthen the validity of the model towards identification of the 1221~nm emitter.

\subsection{Methodology}
First principles plane-wave supercell density functional theory (DFT) calculations are carried out to study the SiV$_2$:H$^{(-)}$ defect in diamond as implemented in the plane-wave \textsc{vasp} \textit{ab initio} code~\cite{Kresse:PRB1996}. The usual projector augmented wave (PAW) method~\cite{Blochl:PRB1994,Blochl:PRB2000} was applied on carbon, silicon and hydrogen atoms with a plane wave cutoff of 370~eV. We employed the hybrid Heyd-Scuseria-Ernzerhof (HSE) DFT functional~\cite{Heyd03, Krukau06} to determine the electronic structure and optical response of the SiV$_2$:H$^{(-)}$ defect. We determined the optically excited state of SiV$_2$:H$^{(-)}$ by means of the $\Delta$SCF method which involves electron-hole interaction and relaxation of ions upon excitation~\cite{Gali2009PRL}. 

\subsection{Level structure of SiV$_2$:H$^{(-)}$}
\begin{figure}[h!]
  \centering
  \includegraphics[width = 105mm]{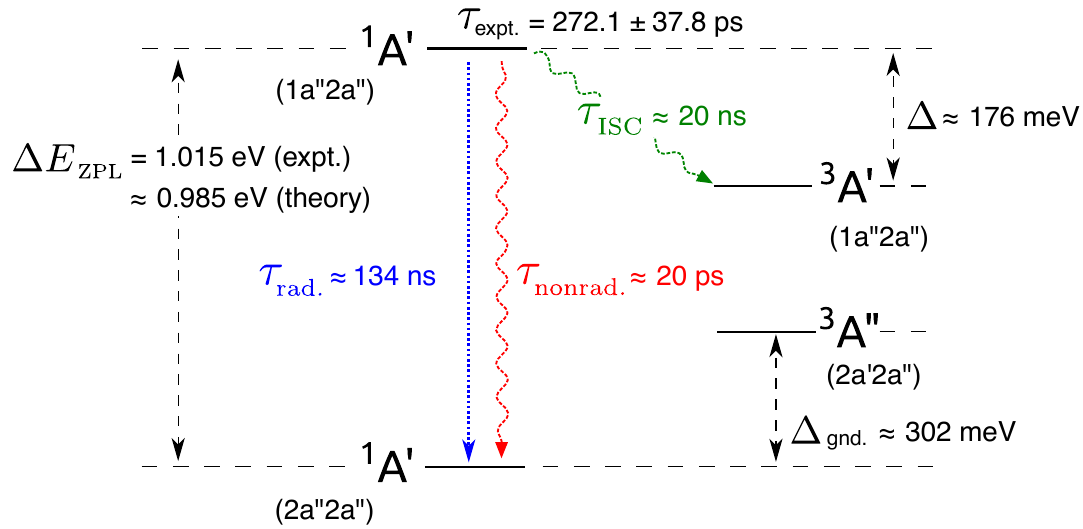}
  \caption{Level structure of the SiV$_2$:H$^{(-)}$  system. We show the symmetry of the unoccupied orbitals in the parentheses for each multiplets that characterize the two-particle wavefunctions.}
  \label{FigSM:levels}
\end{figure}

The level structure was already reported in our previous theoretical study~\cite{thiering2015complexes}. We briefly summarize it here: two deep levels ($\mathrm{2a^\prime}$ and $\mathrm{1a^{\prime\prime}}$) are fully occupied by four electrons whereas a level above ($\mathrm{2a^{\prime\prime}}$) is empty which are all localized to the dangling bonds of the defect. The optical transition between $2a^{\prime\prime}\leftrightarrow1a^{\prime\prime}$ levels can be described by promoting an electron from an orbital localized on a vacancy near the Si ion to the neighbor vacancy. In our previous study, we found two triplet levels between the singlet ground state and excited state (Fig.~\ref{FigSM:levels}).

We improved the method for calculating the associated ZPL energy with respect to that of the previous study by enabling further relaxation of the Kohn-Sham wavefunctions in the $\Delta$SCF procedure which yields 0.99~eV that agrees well with the experimental data (1.015~eV).

\subsection{Lifetime of the excited state}

The observed optical lifetime is
$\tau_\text{expt.} = 272.1 \pm 37.8 \;\text{ps}$. The optical lifetime involves the radiative (rad.) and nonradiative lifetimes. The nonradiative lifetime involves two components, the intersystem crossing (ISC) through the triplet shelving states and the direct one (nonrad.) from the singlet excited state to the ground state. The theoretical optical lifetime which $\tau_\text{theory}$  can be expressed as 
$\tau_\text{theory}^{-1} = \tau_\text{rad.}^{-1} + \tau_\text{nonrad.}^{-1}  + \tau_\text{ISC}^{-1}$.

\subsubsection*{Radiative lifetime}

We determine the radiative lifetime by the following \begin{equation}
\tau_{\mathrm{rad}.}=\frac{3\pi\varepsilon_{0}}{n\omega^{3}|\mu|^{2}}
\label{eqSM:rad}
\end{equation}
equation which results in $\tau_\text{rad.}\approx134\;\text{ns}$. We approximate the optical transition dipole moment $\mu$ by calculating the optical dipole moment between the occupied $1a^{\prime}$ and empty $2a^{\prime\prime}$ single particle orbitals. Additionally, $n=2.42$ is the refractive index of diamond, $\omega$ is the frequency of the emitted photon, and $\varepsilon_0$ is the vacuum permittivity. The reason behind the relatively long lifetime is that the overlap between the ground state and excited state wavefunctions is relatively small, see $2a^{\prime}$ and $2a^{\prime\prime}$ orbitals of $\mathrm{SiV_2:H}$ in Fig.~10 of Ref.~\onlinecite{thiering2015complexes} for comparison.

\subsubsection*{Intersystem crossing lifetime}

\begin{figure}[h!]
  \centering
  \includegraphics[width = 86mm]{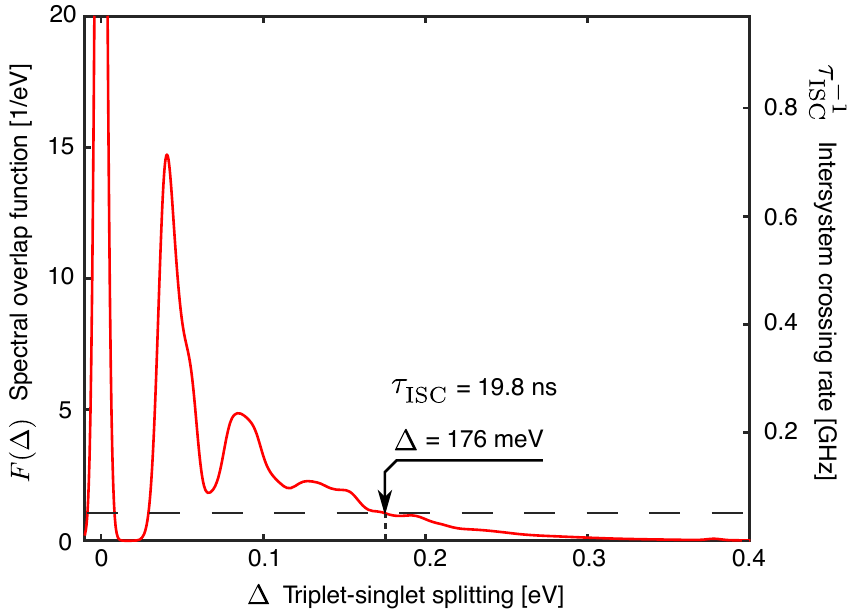}
  \caption{Intersystem crossing rate by means of the Huang-Rhys or Franck-Condon approximation as evaluated by Eq.~\eqref{eqSM:ISC}.}
  \label{FigSM:ISC}
\end{figure}

The ISC lifetime can be calculated by means the Condon approximation \cite{Marian_2011, Goldman_PRB, Goldman_PRL, Thiering_SOC} as the following equation \begin{equation}
\tau_{\text{ISC}}^{-1}=\frac{2\pi}{\hbar}\sum_{m_s=+1,-1,0}\left|\langle^{3}A_{m_s}^{\prime}|\hat{W}_{\mathrm{SOC}}|^{1}A^{\prime}\rangle\right|^{2}F(\Delta) \text{,}
\label{eqSM:ISC}
\end{equation}
which contains the spin-orbit coupling between the singlet and triplet excited states that are separated by $\Delta\approx176$~meV energy. Finally, $F(\Delta)$ is the spectral overlap function of the phonons associated with the two electronic states. The spin-orbit matrix elements between the occupied and empty single particle states associated with the singlet and triplet many-body states are calculated within the scalar relativistic approximation~\cite{Steiner2016}. We determined the spin-orbit matrix elements by means of the PBE (Ref. \cite{PBE}) DFT functional that we will discuss in the next paragraph. The calculations finally leads (Fig.~\ref{FigSM:ISC}) to $\tau_\text{ISC}\approx 20 \;\text{ns}$ where we assume that the transition can lead to $m_s=\pm1$ spin states of the excited triplet. We note that $m_s=0$ is exempt from coupling in Eq.~\eqref{eqSM:ISC} since there will be no corresponding term in $\hat{W}_{\text{SOC}}$ in our level of approximation see the text between Eqs.~\eqref{eq:-2} and ~\eqref{eq:-2} for details. However, traditional DFT calculations can only predict the electronic structure only approximately thus there is a $\sim 0.1$~eV inaccuracy in $\Delta$. Therefore, there might be an order of magnitude error in $\tau_{\text{ISC}}^{-1}$ since $F(\Delta)$ inherits the uncertainty of $\Delta$ exponentially. There is an additional inaccuracy in the spin-orbit coupling; thus the present result is a rough estimate.

\subsubsection*{Spin-orbit coupling between the excited singlet and triplet}
First, we check the spin-orbit coupling within the multiplets. That
is, $|^{1}A^{\prime}\rangle$ belongs to $A^{\prime}$ irreducible
representation and $|^{3}A^{\prime}\rangle$ splits into $\underset{m_s=\{+,-\}}{\underbrace{2A^{\prime}}}\oplus\underset{m_s=0}{\underbrace{A^{\prime\prime}}}$
irreducible representations. We set the spin quantization axis along $z$ direction that
is perpendicular to the mirror axis of $C_{s}$ point group. Therefore, $|^{1}A^{\prime}\rangle$ can only couple to $m_s=\pm1$
part of $|^{3}A^{\prime}\rangle$. Therefore we need to compute the
following coupling,
\begin{equation}
\frac{1}{2}\left(\langle1a^{\prime\prime}2a^{\prime\prime}|+\langle1a^{\prime\prime}2a^{\prime\prime}|\right)\left(\langle\uparrow\downarrow|-\langle\downarrow\uparrow|\right)\hat{W}_{\mathrm{SOC}}\frac{1}{\sqrt{2}}\left(|1a^{\prime\prime}2a^{\prime\prime}\rangle-|1a^{\prime\prime}2a^{\prime\prime}\rangle\right)|\uparrow\uparrow\rangle\text{,}\label{eq:}
\end{equation}
where $\hat{W}_{\mathrm{SOC}}$ is the single particle spin-orbit
(SOC) operator,

\begin{equation}
\hat{W}_{\mathrm{SOC}}=\sum_{\alpha}^{x,y,z}\lambda_{\alpha}\hat{l}_{\alpha}\hat{s}_{\alpha}\:\:\text{.}\label{eq:-1}
\end{equation}
However, it is more convenient to express the $\hat{l}_{\alpha}$
orbital operators by projectors of $|1a^{\prime\prime}\rangle$ and
$|2a^{\prime\prime}\rangle$ orbitals,

\begin{equation}
\begin{split}\hat{W}_{\mathrm{SOC}}=\sum_{\alpha}^{x,y} \Big(& \lambda_{1\alpha}|1a^{\prime\prime}\rangle\langle1a^{\prime\prime}|\hat{s}_{\alpha}+\lambda_{2\alpha}|2a^{\prime\prime}\rangle\langle2a^{\prime\prime}|\hat{s}_{\alpha}\\
+ & \lambda_{3\alpha}|1a^{\prime\prime}\rangle\langle2a^{\prime\prime}|\hat{s}_{\alpha}+\lambda_{4\alpha}|2a^{\prime\prime}\rangle\langle1a^{\prime\prime}|\hat{s}_{\alpha}\Big)\:\:\text{.}
\end{split}
\label{eq:-2}
\end{equation}

Note that the $\hat{s}_{z}$ term is missing because it transforms as
an $A^{\prime\prime}$ and all $|...\rangle\langle...|$ projectors
transform as $A^{\prime}$. The $\hat{W}_{\mathrm{SOC}}$ Hamiltonian
can contain only totally symmetric $A^{\prime}$ terms but $A^{\prime}\otimes A^{\prime\prime}=A^{\prime\prime}$
is not such a term. Therefore, only the $\hat{s}_{x}$ and $\hat{s}_{y}$
spin flipping terms can exist in the single particle spin-orbital picture,

\begin{equation}
\begin{split}\hat{W}_{\mathrm{SOC}}= & \lambda_{1}|1a_{\uparrow}^{\prime\prime}\rangle\langle1a_{\downarrow}^{\prime\prime}|+\lambda_{2}|2a_{\uparrow}^{\prime\prime}\rangle\langle2a_{\downarrow}^{\prime\prime}|+\text{h.c.}\\
+ & \lambda_{3}|1a_{\uparrow}^{\prime\prime}\rangle\langle2a_{\downarrow}^{\prime\prime}|+\lambda_{4}|2a_{\uparrow}^{\prime\prime}\rangle\langle1a_{\downarrow}^{\prime\prime}|+\text{h.c.}\:\:\:\text{.}
\end{split}
\label{eq:-3}
\end{equation}

According to our \textit{ab initio} results $\lambda_{1}$,$\lambda_{2}$ terms
are negligible (smaller than 1~$\mu$eV) and $|\lambda_{3}|=|\lambda_{4}|\approx0.133\:\text{meV}$. We determined this value between the spin-orbit coupling matrix element between the occupied $1a^\prime$ and empty $2a^\prime$ single particle levels in the ground state singlet. 
However, one may still check that $\langle^{3}A^{\prime}|\hat{W}_{\mathrm{SOC}}|^{1}A^{\prime}\rangle=0$,
e.g., the effect of SOC is zero at first order of approximation. In order to have the first non-vanishing term we do the following. 
Within the singlet and triplet states the orbitals can be different in actual DFT calculations,

\begin{equation}
|1a_{\uparrow}^{\prime\prime}+c\times2a_{\uparrow}^{\prime\prime};2a_{\downarrow}^{\prime\prime}+d\times1a_{\downarrow}^{\prime\prime}\rangle=\sqrt{1-c^{2}-d^{2}-c^{2}d^{2}}|1a_{\uparrow}^{\prime\prime}2a_{\downarrow}^{\prime\prime}\rangle+c|2a_{\uparrow}^{\prime\prime}2a_{\downarrow}^{\prime\prime}\rangle+d|1a_{\uparrow}^{\prime\prime}1a_{\downarrow}^{\prime}\rangle+cd|2a_{\uparrow}^{\prime\prime}1a_{\downarrow}^{\prime\prime}\rangle
\label{eq:-4}
\end{equation}
that leads to dynamic correlation between the ground and excited state
singlets. We calculated the overlap between the triplet and singlet
occupied $1a^{\prime\prime}$ and $2a^{\prime\prime}$ orbitals, thus
$d<0.01$ and $c\approx0.378$. Therefore,
\begin{equation}
\langle^{3}A^{\prime}|\hat{W}_{\mathrm{SOC}}|^{1}A^{\prime}\rangle\approx\langle2a_{\uparrow}^{\prime\prime}2a_{\downarrow}^{\prime\prime}|_{\mathcal{A}}c\hat{W}_{\mathrm{SOC}}|1a_{\downarrow}^{\prime\prime}2a_{\downarrow}^{\prime\prime}\rangle_{\mathcal{A}}=c\lambda_{4}^{*}=0.0504\:\text{meV}
\text{,}
\label{eq:-5}
\end{equation}
that we will use in Eq.~\eqref{eqSM:ISC}. We note that $\mathcal{A}$ refers to antisymmetrized wavefunctions $|ab\rangle_{\mathcal{A}}=(|ab\rangle-|ab\rangle)/\sqrt{2}$. We used this spin-orbit coupling parameter in Eq.~\eqref{eqSM:ISC} that exists only the correlated nature of the singlet levels.
\subsubsection*{Phonon assisted direct nonradiative relaxation}

\begin{figure}[h!]
  \centering
  \includegraphics[width = 130mm]{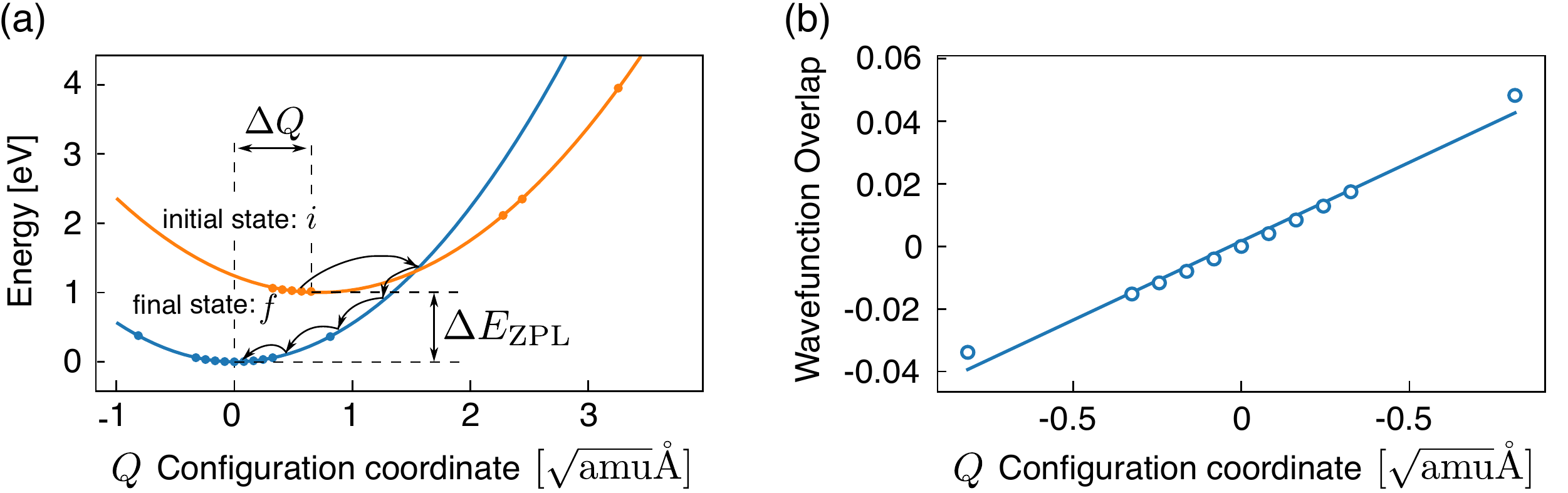}
  \caption{(a) Adiabatic potential energy surface of SiV$_2$:H$^{(-)}$. We set $\Delta E_{\text{ZPL}} = 1.015$~eV since the experimentally measured ZPL is more reliable. Upon electronic excitation the equilibrium of position is displaced by $\Delta Q$ = 0.65133~$\sqrt{\text{amu}} \text{\AA}$. (b) Fitting on wavefunction overlaps $\langle\psi_{i}(0)|\psi_{f}(Q)\rangle$ between the ground (\textit{final}) occupied electronic orbital and excited (\textit{initial}) empty electronic orbital to determine $W_{if}$. However, $W_{if}$ is multiplied by an additional energy factor $\hat{H}=(\varepsilon_{\text{KS}}^{f}-\varepsilon_{\text{KS}}^{i})|1a^{\prime\prime}\rangle\langle2a^{\prime\prime}|+\text{h.c.}$ where $(\varepsilon_{\text{KS}}^{f}-\varepsilon_{\text{KS}}^{i})=1.89$ eV is the difference that of the occupied and empty pair of Kohn-Sham levels participating in the nonradiative process.}
  \label{figSM:nonrad}
\end{figure}

The nonradiative transition is calculated based on NONRAD code~\cite{alkauskas_first-principles_2014, turiansky_nonrad_2021}, as discussed in Supplementary Note 9 in Ref.~\cite{Li_2022}. Therefore we evaluate the following expression \begin{equation}
{\tau_{\text{nonrad.}}}^{-1}=\frac{2\pi}{\hbar}W_{if}^{2}\sum_{m,n}p_{i,m}\left|\langle\chi_{i,m}|\hat{Q}-Q_{0}|\chi_{f,n}\rangle\right|^{2}\delta(m\hbar\omega_{i}-n\hbar\omega_{f}+\Delta E_{\text{ZPL}})
\label{eqSM:nonrad}
\end{equation} to evaluate the phonon assisted nonradiative rates. Here, $W_{if}=\langle\psi_{i}|\partial_{Q}\hat{H}|\psi_{f}\rangle = 0.09541$~$\text{eV} /(\sqrt{\text{amu}} \text{\AA}) $ is linear electron-phonon matrix element that we determine by fitting electronic wavefunction overlaps respect to $Q$ configurational coordinate that connects the excited state and ground state geometries (Fig.~\ref{figSM:nonrad}(b)). $p_{i,m}$ corresponds to the thermal population of $i$ (initial) state's $m^\text{th}$ vibration level and $f$ corresponds to the final state. The phonon matrix $\langle\chi_{i,m}|\hat{Q}-Q_{0}|\chi_{f,n}\rangle$ sums up the harmonic oscillator wave functions that enter the nonradiative recombination process. $\hat{Q}$ is the coordinate operator that spans the harmonic oscillator's creation and annihilation operators, $\hat{Q}=\sqrt{\hbar/2\omega_{i/f}}(\hat{a}_{i/f}^{\dagger}+\hat{a}_{i/f})$. $Q_0$ is the atomic configuration used as the starting point for the perturbative expansion. Finally, we determine $\omega_i=61.9$~meV and $\hbar\omega_f=68.4$~meV frequencies by fitting the adiabatic potential energy surface (Fig.~\ref{figSM:nonrad}(a)). 

According to our results, the nonradiative lifetime is $\tau_\text{nonrad.} = 19.6\;\text{ps}$ between $T=0$ and $100$~K. Additionally, we predict that at elevated temperatures the rate is significantly faster: $15.2$~ps at 200~K and $8.6$~ps at 300~K due to thermally activated phonons. The calculated nonradiative lifetime is an order of magnitude shorter than the lifetime observed by transient absorption. We note that the nonradiative lifetime is very sensitive to the geometry change. As the singlet excited state is a multiplet in which the constraint DFT method has limited accuracy, we cannot disregard that the calculated nonradiative lifetime is underestimated. Even taking this discrepancy into account, these calculations imply that if the measured O-band emitter is indeed the SiV$_2$:H$^{(-)}$ defect then its decay from the excited state is dominated by direct nonradiative relaxation process.

\bibliography{TelecomEmitter}